\documentclass[reprint,aps,superscriptaddress]{revtex4-1}
\usepackage{graphicx}
\usepackage{amsmath}

\begin{document}
\title{$J_1$-$J_2$ square lattice antiferromagnetism in the orbitally quenched insulator $\mathbf{MoOPO_4}$}

\author{L. Yang}
\affiliation{Laboratory for Quantum Magnetism, Institute of Physics, Ecole Polytechnique F\'{e}derale de Lausanne (EPFL), CH-1015 Lausanne, Switzerland}
\affiliation{Laboratory of Physics of Complex Matter, Institute of Physics, Ecole Polytechnique F\'{e}derale de Lausanne (EPFL), CH-1015 Lausanne, Switzerland}
\author{M. Jeong}
\email{minki.jeong@gmail.com}
\affiliation{Laboratory for Quantum Magnetism, Institute of Physics, Ecole Polytechnique F\'{e}derale de Lausanne (EPFL), CH-1015 Lausanne, Switzerland}
\author{P. Babkevich}
\affiliation{Laboratory for Quantum Magnetism, Institute of Physics, Ecole Polytechnique F\'{e}derale de Lausanne (EPFL), CH-1015 Lausanne, Switzerland}
\author{Vamshi M. Katukuri}
\affiliation{Chair of Computational Condensed Matter Physics, Institute of Physics, Ecole Polytechnique ~F\'{e}derale de Lausanne (EPFL), CH-1015 Lausanne, Switzerland}
\author{B.~N\'{a}fr\'{a}di}
\affiliation{Laboratory of Physics of Complex Matter, Institute of Physics, Ecole Polytechnique F\'{e}derale de Lausanne (EPFL), CH-1015 Lausanne, Switzerland}
\author{N. E. Shaik}
\affiliation{Laboratory for Quantum Magnetism, Institute of Physics, Ecole Polytechnique F\'{e}derale de Lausanne (EPFL), CH-1015 Lausanne, Switzerland}
\author{A. Magrez}
\affiliation{Crystal Growth Facility, Institute of Physics, Ecole Polytechnique F\'{e}derale de Lausanne (EPFL), CH-1015 Lausanne, Switzerland}
\author{H. Berger}
\affiliation{Crystal Growth Facility, Institute of Physics, Ecole Polytechnique F\'{e}derale de Lausanne (EPFL), CH-1015 Lausanne, Switzerland}
\author{J. Schefer}
\affiliation{Laboratory for Neutron Scattering and Imaging, Paul Scherrer Institute (PSI), CH-5232 Villigen, Switzerland}
\author{E. Ressouche}
\affiliation{Univ. Grenoble Alpes, CEA, INAC, MEM, F-38000 Grenoble}
\author{M. Kriener}
\affiliation{RIKEN Center for Emergent Matter Science (CEMS), Wako, Saitama 351-0198, Japan}
\author{I.~\v Zivkovi\'c}
\affiliation{Laboratory for Quantum Magnetism, Institute of Physics, Ecole Polytechnique F\'{e}derale de Lausanne (EPFL), CH-1015 Lausanne, Switzerland}
\author{O. V. Yazyev}
\affiliation{Chair of Computational Condensed Matter Physics, Institute of Physics, Ecole Polytechnique ~F\'{e}derale de Lausanne (EPFL), CH-1015 Lausanne, Switzerland}
\author{L. Forr\'{o}}
\affiliation{Laboratory of Physics of Complex Matter, Institute of Physics, Ecole Polytechnique F\'{e}derale de Lausanne (EPFL), CH-1015 Lausanne, Switzerland}
\author{H. M. R\o nnow}
\email{henrik.ronnow@epfl.ch}
\affiliation{Laboratory for Quantum Magnetism, Institute of Physics, Ecole Polytechnique F\'{e}derale de Lausanne (EPFL), CH-1015 Lausanne, Switzerland}
\affiliation{RIKEN Center for Emergent Matter Science (CEMS), Wako, Saitama 351-0198, Japan}

\begin{abstract}
We report magnetic and thermodynamic properties of a $4d^1$ (Mo$^{5+}$) magnetic insulator $\mathrm{MoOPO_4}$ single crystal, which realizes a $J_1$-$J_2$ Heisenberg spin-$1/2$ model on a stacked square lattice. The specific-heat measurements show a magnetic transition at 16 K which is also confirmed by magnetic susceptibility, ESR, and neutron diffraction measurements. Magnetic entropy deduced from the specific heat corresponds to a two-level degree of freedom per Mo$^{5+}$ ion, and the effective moment from the susceptibility corresponds to the spin-only value. Using {\it ab initio} quantum chemistry calculations we demonstrate that the Mo$^{5+}$ ion hosts a purely spin-$1/2$ magnetic moment, indicating negligible effects of spin-orbit interaction. The quenched orbital moments originate from the large displacement of Mo ions inside the MoO$_6$ octahedra along the apical direction. The ground state is shown by neutron diffraction to support a collinear N\'eel-type magnetic order, and a spin-flop transition is observed around an applied magnetic field of 3.5 T. The magnetic phase diagram is reproduced by a mean-field calculation assuming a small easy-axis anisotropy in the exchange interactions. Our results suggest $4d$ molybdates as an alternative playground to search for model quantum magnets.
\end{abstract}

\maketitle

\section{Introduction}

The $4d$ transition-metal oxides naturally bridge the two different regimes of the strongly correlated $3d$ compounds and the $5d$ compounds with strong spin-orbit coupling (SOC) \cite{Khomskii}. To what extent the $4d$ compounds represent the either regime or display original properties are largely open questions of current interest \cite{Witczak13ARCMP}. Most notably, for instance, it is intriguing that seemingly similar $\mathrm{Ca_2RuO_4}$ and $\mathrm{Sr_2RuO_4}$ display totally different behavior: the former is a Mott insulator \cite{Nakatsuji97JPSJ, Cao97PRB, Nakatsuji00PRL, Fatuzzo15PRB} while the latter is a metal and becomes superconducting at low temperature \cite{Nakatsuji00PRL, Maeno94Nat, ishida1998spin, Fatuzzo15PRB}. Despite a large interest, however, purely $4d$ quantum (spin-$1/2$) magnets are rather rare \cite{deVries10PRL, Aharen10PRB, Clark14PRL, Ishikawa} as the electronic structure is often complicated by the presence of other types of, e.g., $3d$ or $4f$ magnetic orbitals \cite{RevModPhys.82.53}.

\begin{figure}[!ht]
	\centering
	\includegraphics[width=0.45\textwidth]{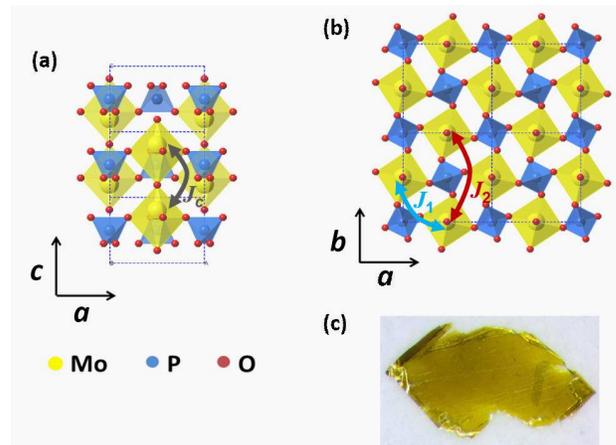}
	\caption{Crystal structure of $\mathrm{MoOPO_4}$ projected onto (a) the $ac$ planes showing a chain-like arrangement of $\mathrm{MoO_6}$ octahedra (yellow) and (b) the $ab$ planes showing the coupling between the chains via $\mathrm{PO_4}$ tetrahedra (blue). Dashed lines represent the unit cells. Possible in-plane ($J_1$ and $J_2$) and out-of-plane ($J_c$) exchange couplings are also shown (c) Photograph of a representative single crystal.}\label{crystal}
\end{figure}

Among a few known $4d^1$ magnets \cite{deVries10PRL, Clark14PRL, Ishikawa} a molybdenium phosphate $\mathrm{MoOPO_4}$ is reported~\cite{Kierkegaard64Acta}. The $\mathrm{MoO_6}$ octahedra with Mo$^{5+}$ ions are corner shared to form a chain along the crystallographic $c$ axis of the tetragonal structure (Fig.~\ref{crystal}(a)), and these chains are further coupled next to each other via corner sharing $\mathrm{PO_4}$ tetrahedra (Fig.~\ref{crystal}(b)) \cite{Kierkegaard64Acta, Canadell97ChemM}. Previous susceptibility data on a powder sample of $\mathrm{MoOPO_4}$ shows a Curie-Weiss behavior with antiferromagnetic $\Theta_{CW}=-8$~K and a magnetic transition at 18 K \cite{Lezama90SSC}. $^{31}$P NMR on a powder evidences a substantial exchange through the $\mathrm{PO_4}$ tetrahedra and a sharp powder ESR line infers a rather isotropic $g$ factor \cite{Lezama90SSC}. However, so far there have not been any studies on the magnetic structure in the ordered state nor magnetic properties of a single crystal. Moreover, any discussion on possible interplay between the crystal electric field and SOC is absent. 

Here we report the magnetic and thermodynamic properties of a $\mathrm{MoOPO_4}$ single crystal using specific heat, susceptibility, magnetization, ESR, and neutron diffraction experiments. We also elucidate the electronic states and magnetic aspects in light of SOC and crystal field effects, with the help of {\it ab initio} quantum chemistry calculations. 

\section{Experimental Details}

High-quality single crystals of $\mathrm{MoOPO_4}$ were grown following the procedure described in Ref.~\cite{Kierkegaard64Acta}. $\mathrm{H_2MoO_4}$ was mixed with concentrated phosphoric acid and heated up to 1000 $^{\circ}$C for reaction in an open platinum crucible. After being cooled to room temperature, the resulting dark-blue solid was dissolved in a large amount of hot water. The yellow transparent crystals were obtained in a plate-like shape (Fig.~\ref{crystal}(c)). Large-sized crystals have a typical dimension of $3\times 2\times 0.4$ mm$^3$ with the $c$ axis normal to the plate. The crystal belongs to the space group of $P4/n$ with the lattice parameters of $a = b = 6.2044$ {\AA} and $c = 4.3003$ {\AA}, obtained by a single-crystal x-ray diffraction, in agreement with Ref.~\cite{Kierkegaard64Acta}.

Specific heat was measured using a physical properties measurement system (PPMS, Quantum Design, Inc.), and magnetization was measured using a magnetic properties measurement system SQUID (MPMS, Quantum Design, Inc.). ESR measurements were performed using a Bruker X-band spectrometer with a TE$_{102}$ resonant cavity around 9.4 GHz. Neutron diffraction experiments were performed on TRICS and D23 beamlines at PSI and ILL, respectively. An incident neutron wavelength of 2.3109 {\AA} was employed.

\section{Results}
\subsection{Specific heat}

\begin{figure}
	\centering
	\includegraphics[width=0.45\textwidth]{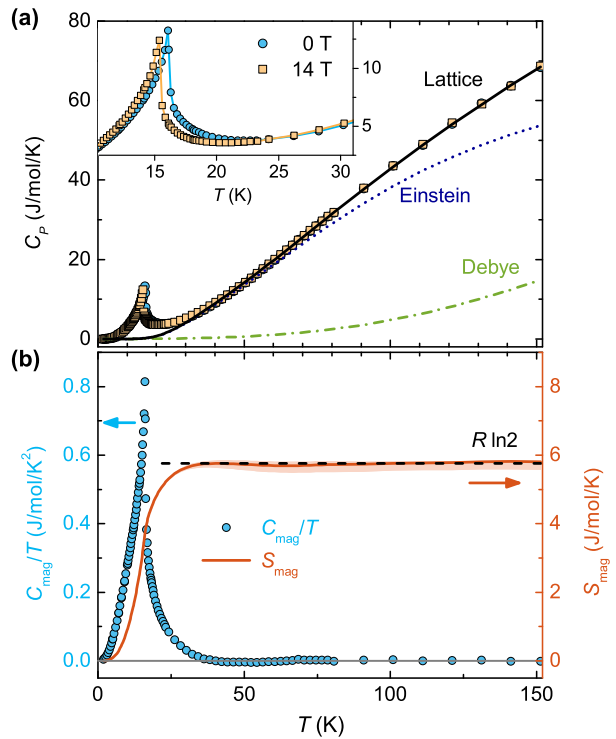}
	\caption{(a) Specific heat $C_p$ as a function of temperature in zero field (circle) and in 14 T (square). Solid line represents the best fit of the simulated lattice contribution using the Debye (dash-dotted line) and Einstein (dotted line) terms. Inset provides an enlarged view of the low temperature region. (b) Left axis: magnetic part of the specific heat, $C_\mathrm{mag}$, divided by temperature (circle). Right axis: solid line is the entropy calculated from the $C_\mathrm{mag}$.}\label{Cp}
\end{figure}

Figure~\ref{Cp}(a) shows the specific heat $C_p$ measured from 2 to 150 K in zero field and in a magnetic field of 14 T. The $C_p$ above 25 K for the both fields is essentally the same, increasing monotonically with increasing temperature. In zero field a pronounced peak is found at 16.1 K, while the peak is shifted to a slightly lower temperature of 15.4 K in 14 T. These peaks correspond to a transition into a magnetically long-range-ordered phase, as evidenced by other experimental measurements discussed in later sections.

\begin{figure*}[!th]
	\centering
	\includegraphics[width=1\textwidth]{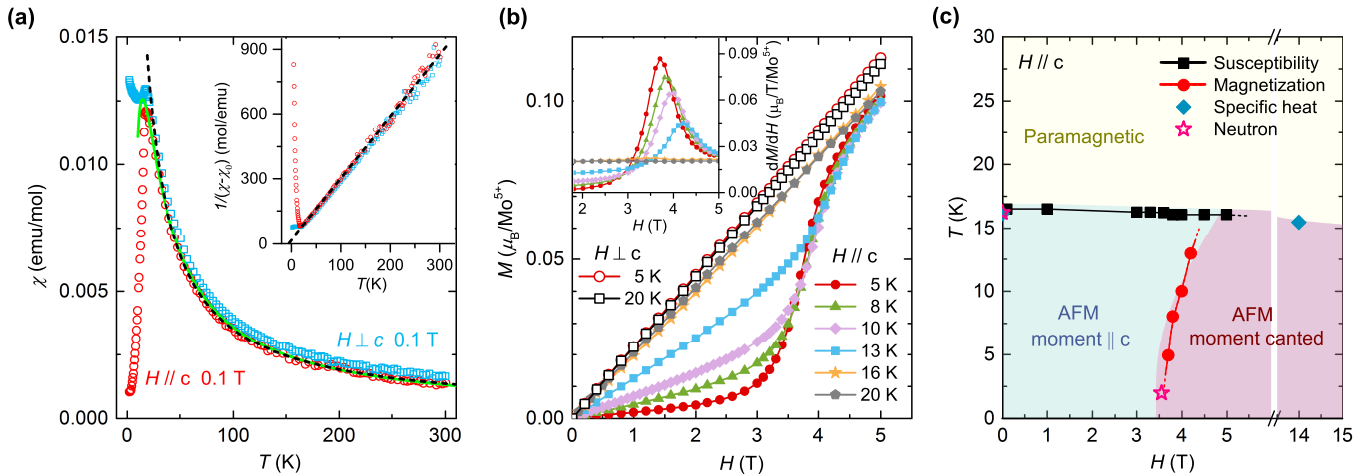}
	\caption{(a) DC magnetic susceptibility $\chi(T)$ in a field of $H=0.1$ T applied parallel (circle) and perpendicular (square) to the $c$ axis. Dashed line represents the Curie-Weiss fit for $H\parallel c$, and solid line high temperature series expansion using Pad\'e approximant, respectively (see the text). Inset shows the inverse susceptibility $ 1/(\chi-\chi_0) $ against temperature where the dashed line is a fit to the Curie-Weiss formula. (b) Isothermal magnetization $M(H)$ for $H\parallel c$ (filled symbol) and $H\perp c$ (open symbol) at several different temperatures. Inset plots the field derivative $dM/dH$ versus $H$ for $H\parallel c$. (c) Magnetic phase diagram from the susceptibility (square), specific heat (diamond), magnetization (circle) and neutron diffraction (star) data. Lines are guides to the eye. Colored background represents the result from the mean field calculations (see text).}\label{magnetic}
\end{figure*}

In order to extract the magnetic part of the specific heat, $C_\mathrm{mag}$, and to deduce the corresponding entropy $S_\mathrm{mag}$, we simulate the lattice contribution from the high temperature data by taking into account the Debye and Einstein contributions. We fit the $C_p$ data above 30 K by a lattice-only model $C_p = C_D + \sum_{i}C_{E,i}$, where $C_{D}$ and $C_{E,i}$ represent the Debye and Einstein terms, respectively. The Debye term is expressed as  
\begin{equation}
C_{D} = 9 n_D R \left(\frac{T}{\Theta_D}\right)^3 \int_{0}^{\Theta_D/T} \frac{x^4 e^x}{\left(e^x-1\right)^2}dx,
\end{equation}
and the Einstein term as
\begin{equation}
C_{E} =3 n_E R \frac{y^2e^{y}}{\left(e^{y}-1\right)^2},\, y\equiv \Theta_E/T,
\end{equation}
where $R$ denotes the gas constant, $\Theta_D$ and $\Theta_E$ are the Debye and Einstein temperatures, and $n_D$ and $n_E$ are the numbers of the corresponding modes, respectively; the sum $n_D+n_E$ is the total number of atoms per formula unit. The best fit for the zero-field, using one Debye and two Einstein terms, yields the characteristic temperatures $\Theta_{D}$ = 1177 K, $\Theta_{E, 1} = 372$ K, and $\Theta_{E, 2} = 154$ K, and the numbers $n_{D}$ = 4, $n_{E, 1}$ = 2, $n_{E, 2}$ = 1. Solid line in Fig.~\ref{Cp}(a) is the best fit result for the total lattice contribution while dash-dotted and dashed lines are the corresponding Debye and Einstein contributions, respectively. While the parameters in the phonon fit may not be directly physical, they provide a parametrisation of the lattice contribution to the specific heat, which can be substracted to estimate the magnetic specific heat.

Figure~\ref{Cp}(b) shows the resulting $C_\mathrm{mag}$ in zero field (circle, left axis) obtained by subtracting the lattice contribution from the measured $C_p$. Solid line in Fig. \ref{Cp}(b) plots the $S_\mathrm{mag}(T)$ obtained by integrating $C_\mathrm{mag}/T$ over temperature (right axis). $S_\mathrm{mag}(T)$ is found to reach and stay at $R\ln 2$ at high temperatures, indicating two-level degrees of freedom. The thin colored band in Fig.~\ref{Cp}(b) represents the entropy range obtained when fitting the $C_p$ data by varying the lower bound of temperature between 25 K to 35 K, to confirm the little dependence of the result on the chosen fit range. The similar analysis for the 14 T data (not shown) indicates negligible field effects.

\subsection{Susceptibility and magnetization}

Figure \ref{magnetic}(a) shows the DC magnetic susceptibility $\chi=M/H$, where $M$ is magnetization, in a field of $H=0.1$ T applied parallel and perpendicular to the $c$ axis. For both cases, $\chi(T)$ shows almost identical behavior from 300 K down to 20 K. However, for $H\parallel c$ the $\chi(T)$ exhibits a sharp drop toward zero as temperature is decreased across 17 K, while the one for $H\perp c$ remains only weakly temperature dependent. This is indicative of antiferromagnetic transition where the ordered moments at low temperatures are collinear to each other, and parallel to the $c$ axis.

The nearly isotropic, high-temperature part of $\chi(T)$ could be well fit by the Curie-Weiss formula, $\chi(T)=C/(T-\Theta_{CW})+\chi_0$, where $\Theta_{CW}$ is the Curie-Weiss temperature and $\chi_0$ a temperature-independent diamagnetic and background term. The best fit yields the effective moment $\mu_\mathrm{eff}=1.67(1)\, \mu_\mathrm{B}$ per Mo$^{5+}$ ion, $\Theta_{CW} = -6(1)$ K, and $\chi_0 = 2.2(1) \times$ $ 10^{-4}$ emu/mol for $H\parallel c$, and $\mu_\mathrm{eff} = 1.69(1)\, \mu_\mathrm{B}$, $\Theta_{CW} = -4(1)$ K, and $\chi_0 = 4.6(1) \times 10^{-4}$ emu/mol for $H\perp c$. The best fit for $H\parallel c$ is shown as dashed line in Fig.~\ref{magnetic}(a). The negative $\Theta_{CW}$ indicates that antiferromagnetic interactions are dominant. The effective moments indicate a spin-only value consistent with the specific-heat results.

\begin{figure*}[!ht] 
	\centering
	\includegraphics[width=1\textwidth]{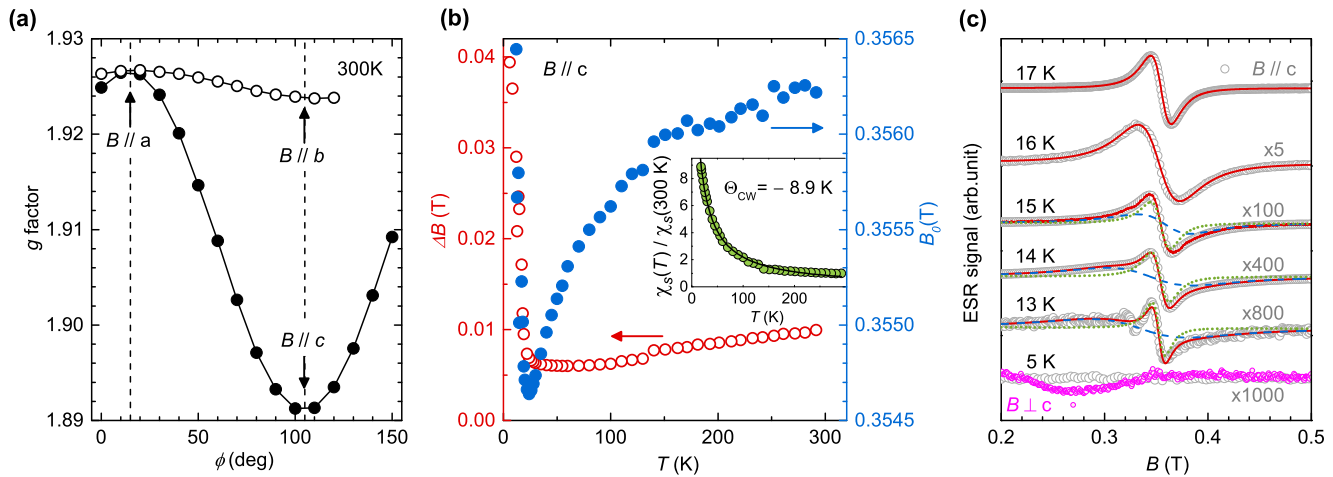}
	\caption{(a) Angular dependence of the $g$ factor at room temperature from the ESR measurements, where filled symbols are for the field orientation varied on the $ac$ plane while open symbols on the $ab$ plane. (b) Resonance field $B_0$ (filled circle, right axis) and line width $\Delta B_0$ (open circle, left axis) of the ESR spectrum as a function of temperature. Inset plots normalized spin susceptibility $\chi_s(T)/\chi_s(300\, \mathrm{K})$ as a function of temperature. (c) Temperature evolution of the spectrum for $B\parallel c$ across the transition (open circles). Solid line is a sum of two contributions from intrinsic (dotted line) and defect (dashed line) susceptibility. At 5 K, the data for $B\perp c$ (filled circle) are overlaid.}\label{ESR}
\end{figure*}

The isothermal magnetization $M(H)$ for $H\parallel c$ and $H\perp c$ at several temperatures are shown in Fig.~\ref{magnetic}(b). At 5 K, $M(H)$ increases slowly with the field $H\parallel c$ up to 3 T, but then sharply increases in a narrow field range of $3-4$~T until it eventually converges to the high temperature $M(H)$ data obtained at 16 K or 20 K. This stepwise increase of $M(H)$ becomes smeared out as temperature is increased. On the other hand, no such stepwise behavior was observed at any temperatures for $H\perp c$. These are typical signatures of a spin-flop transition which occurs when the field is applied along an easy axis, along which the ordered moments align. 

The magnetic phase diagram is thus mapped out by combining the above bulk magnetic and specific-heat results, as shown in Fig.~\ref{magnetic}(c). The antiferromagnetic transition temperatures in different fields are obtained from the peaks in $\chi(T)$ and $C_p(T)$, and the spin-flop transition fields at different temperatures are obtained from the peak positions in the $dM/dH$ versus $H$ plot (inset of Fig.~\ref{magnetic}(b)).

\subsection{Electron spin resonance}

In order to gain a microscopic insight into the magnetic properties, we have performed ESR measurements as a function of field orientation and temperature. Figure~\ref{ESR}(a) plots the obtained room-temperature $g$ factor as the field direction is rotated by $\phi$ in the $ab$ and the $ac$ planes. The $g$ factor in the $ac$ plane shows a $\phi$-variation as large as 2 $\%$ with characteristic $\cos ^2\phi$ angular dependence. On the other hand, the $g$ factor in the $ab$ plane remains essentially constant as expected from the tetragonal symmetry, within the error of 0.08 \% which might have arisen from a slight misorientation of the crystal. We obtain the $g$ factor along the principal axes as $g_a = 1.926(2)$ and $g_c=1.889(2)$. The average value $g = (2g_a + g_c)/3 = 1.913(2)$ agrees with the one previously obtained by powder ESR \cite{Lezama89SSC}. These $g$ values correspond to the effective moments of 1.64 and 1.66~$\mu_B$ for spin-$1/2$, for $H\parallel c$ and $H\perp c$, respectively, which are very close to the effective moment values obtained from the Curie-Weiss fit in the previous section.

For a system with tetragonal symmetry with short distances between the transition metal and ligand ions, one would expect $g_a < g_c$~\cite{book_abragam_bleaney}. However, we find an opposite structure for the $g$ factor in MoOPO$_4$, even though the orbital energy diagram  for the Mo$^{5+}$ ion is expected to be similar to that of tetragonally compressed octrahedron with stabilized $d_{xy}$ orbital (see Fig.~\ref{calc}). As explained in Sect. E, the multi-orbital character of the ground state in MoOPO$_4$ results in the observed $g$ values. 

Figure~\ref{ESR}(b) shows the temperature dependence of the resonance field $B_0$ and the line width $\Delta B_0$ of the ESR spectrum. $B_0$ slowly decreases as temperature is lowered from 300 K down to 24 K, which may be attributed to a lattice contraction. As temperature is further lowered below 24 K, $B_0$ starts increasing sharply, which indicates that a magnetic transition is approached. Similarly, $\Delta B_0$ slowly decreases as temperature is lowered down to 25 K, but then starts broadening significantly as temperature is further lowered down to 15 K due to critical spin fluctuations. Inset of Fig.~\ref{ESR}(b) plots temperature dependence of the local spin susceptibility which is obtained from the spectral area at each temperature normalized by the one at 300 K, $\chi_s(T)/\chi_s(\mathrm{300\, K)}$. The data could be fit to the Curie-Weiss formula with $\Theta_{CW}=-8.9$ K, which is in reasonable agreement with the bulk susceptibility result shown in Fig.~\ref{magnetic}(a). 

Across the transition, the ESR line changes in shape and intensity as shown in Fig.~\ref{ESR}(c). The line sustains a perfect Lorenztian shape down to 16 K. On the other hand, the line below 16 K close to the transition fits better to a sum of two Lorentzian: one corresponds to the intrinisic sample susceptibility while the other may correspond to some defects. Indeed, the ESR signal at the paramagnetic resonance field position below 15 K corresponds to about 0.1 \% concentration of paramagnetic impurities. The response below 15 K represents the summation of the possible defect contribution and the intrinsic susceptibility. The tiny intrinsic response below the transition temperature may represent clusters of which spins remain fluctuating within the ESR time window, which essentially disappears at lower temperatures below 14 K. At 5 K, a broad hump of weak signal is observed around 0.27 T for $\mathrm{B}\parallel c$, which is absent for $B\perp c$. This signal may correspond to an antiferromagnetic resonance.

\subsection{Neutron diffraction}

\begin{figure}[!b]
\centering
\includegraphics[width=0.48\textwidth]{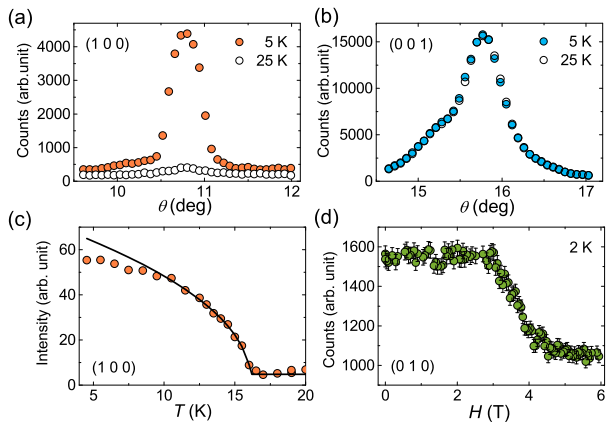}
\caption{Neutron diffraction measurements of rotation scans through (a): (100) and (b): (001) reflections recorded at 5 and 25 K. (c) Temperature evolution of the Bragg peak integrated intensity. The solid line is a power-law fit $I(T)\propto(T_N-T)^{2\beta}$ with the parameters $\beta=0.23$ and $T_N=16.17 \pm 0.06$ K. (d) (010) peak counts as a function of magnetic field parallel to $c$ axis.}\label{ns}
\end{figure}

To determine the microscopic magnetic structure, we have performed neutron diffraction measurements. Magnetic intensity appears at the position of the $k=(100)$ wave vector at 5\,K, as shown in the rotation scan in Fig.~\ref{ns}(a). No appreciable change in scattering is found close to $(001)$ between 5 and 25\,K as shown in Fig.~\ref{ns}(b). A small shoulder of the $(001)$ reflection is likely to originate from a closely oriented secondary grain. A non-zero $(100)$ reflection would be consistent with Mo spins related by a spatial inversion being antiparallel. Due to the dipolar nature of the magnetic interaction, only magnetization perpendicular to the scattering wavevector gives a non-zero structure factor. As no change is observed for the $(001)$ reflection upon cooling below $T_N$, we can conclude that the moments are parallel to the $c$ axis. To verify that this is consistent with the symmetry of the lattice and rule out any other magnetic structures, we utilize BasiReps \cite{Fullprof} and outline the results here.

The magnetic representation is decomposed into six one-dimensional irreducible representations $\Gamma_\nu$ whose resulting basis functions are shown in Table~\ref{tab:basis_vectors}. Examining the results of the irreducible magnetic representations, we find that only $\Gamma_2$ is consistent with our observations. These results are in contrast to the closely related $A$MoO(PO$_4$)Cl ($A=$ K and Rb) materials. Unlike the tilted arrangement of MoO$_6$ octahedra and PO$_4$ tetrahedra in MoOPO$_4$, $A$MoO(PO$_4$)Cl possesses a higher symmetry where the octahedra and tetrahedra are arranged untilted in the $ab$ plane \cite{Borel98}. Powder neutron diffraction measurements on $A$MoO(PO$_4$)Cl reveal an antiferromagnetic structure where Mo moments are instead confined to the $ab$ plane \cite{Ishikawa}.

\begin{table}[!t]
\centering
\begin{tabular}{ c  c  r r}
\hline
\hline
$\nu$ &   & Mo$_1$   & Mo$_2$ \\
\hline
1 & Re & $(0,0,1)$ & $(0,0,1)$ \\
\hline
2 & Re & $(0,0,1)$ & $(0,0,\bar{1})$ \\
\hline
3 & Re & $(1,0,0)$ & $(1,0,0)$ \\
3 & Im & $(0,\bar{1},0)$ & $(0,\bar{1},0)$ \\
\hline
4 & Re & $(1,0,0)$ & $(\bar{1},0,0)$ \\
4 & Im & $(0,\bar{1},0)$ & $(0,1,0)$ \\
\hline
5 & Re & $(1,0,0)$ & $(1,0,0)$ \\
5 & Im & $(0,1,0)$ & $(0,1,0)$ \\
\hline
6 & Re & $(1,0,0)$ & $(\bar{1},0,0)$ \\
6 & Im & $(0,1,0)$ & $(0,\bar{1},0)$ \\
\hline
\hline
\end{tabular}
\caption{Basis functions of irreducible representation $\Gamma_\nu$ for $k=(100)$ separated into real and imaginary components and resolved along the crystallographic axes. The two equivalent Mo$_1$ and Mo$_2$ ions are related by an inversion through the origin.
\label{tab:basis_vectors}}
\end{table}

Figure~\ref{ns}(c) shows the temperature dependence of the (100) Bragg peak integrated intensity. By fitting a powerlaw dependence to the intensity, we find $T_N=16.17 \pm 0.06$ K -- consistent with the magnetization and specific-heat measurements. The order parameter exponent is found to be $\beta=0.23$, corresponding to the 2D XY universality class, However, dedicated measurements with better resolution and separating critical scattering would be needed before any conclusions are drawn from this. In Fig.~\ref{ns}(d) we show the magnetic Bragg peak intensity as a function of applied field along the $c$ axis recorded at 2\,K. Above 3\,T, we find a sharp decrease in intensity which then appears to saturate above 5\,T. The change in the Bragg peak intensity is consistent with a spin-flop transition that is observed in the magnetization measurements shown in Fig.~\ref{magnetic}(b). This corresponds to a tilt of the moments by approximately 35$^\circ$ away from the $c$ axis for the fields above 5 T.

\subsection{Model calculations}

\begin{figure}[!t]
	\centering
	\includegraphics[width=0.4\textwidth]{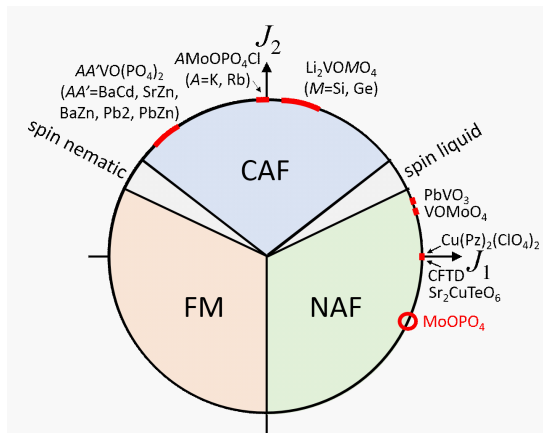}
	\caption{Schematic phase diagram of the spin-$1/2$ $J_1$-$J_2$ Heisenberg model on a square lattice with corresponding compounds~\cite{Ishikawa}. CAF, NAF and FM refer to columnar antiferromagnetic, N\'eel antiferromagnetic and ferromagnetic ground state, respectively.}\label{j1j2}
\end{figure}

In order to gain insight into the magnetic interactions, we have fit the experimental susceptibility shown in Fig.~\ref{magnetic}(a) using a high-temperature series expansion \cite{Schmidt11PRB} assuming a $J_1$-$J_2$ spin-$1/2$ Heisenberg model on a square lattice. The best fit (solid line in Fig.~\ref{magnetic}(a)) returns $J_1=11.4(0.4)$ K and $J_2=-5.2(1.0)$ K corresponding to $J_2/J_1=-0.46$. This ratio supports a collinear N\'eel order for the ground state (see Fig.~\ref{j1j2}) in agreement with the neutron diffraction result. Using the mean-field expression for the Curie-Weiss temperature,
\begin{equation}
\Theta_{CW} = -\frac{S(S+1)}{3k_B}\sum_{i=1,2} z_i J_i,
\end{equation}
where $z_i$ is the number of neighbors for the corresponding couplings (4 both for $J_1$ and $J_2$ in the present case), the high-temperature expansion fit yields $\Theta_{CW}=-6.2$ K, which agrees with the value obtained from the simple Curie-Weiss fit. Next, we simulate the phase diagram using a mean-field calculation. The results are presented by the colored background in Fig.~\ref{magnetic}(c). A slight exchange anisotropy, $\Delta=0.02$, has been introduced in the Hamiltonian, i.e., for a pairwise interaction $\mathcal{H}= J[S^xS^x + S^yS^y + (1+\Delta) S^zS^z]$, to account for the spin-flop transition in a spin-$1/2$ system where single-ion anisotropy is not expected to be present. We note that the mean-field calculation reproduces the temperature dependence of the spin-flop field. From the mean-field expression for the N\'eel temperature, 
\begin{equation}
T_{N} = -\frac{S(S+1)}{3k_B}\sum_{i=1,2} z_i (-1)^i J_i,
\end{equation}
we obtain $T_N = 16.6(1.4)$~K, which is in excellent agreement with the actual value from the experiments.

With strong ferromagnetic second nearest neighbor interactions, MoOPO$_4$ populates a region of the $J_1$-$J_2$ phase diagram, which has so far seen rather few investigations (see Fig.~\ref{j1j2}). In the context of $(\pi,0)$ zone-boundary anomalies on the square lattice, linear spin-wave theory would for MoOPO$_4$ predict a dispersion with significantly higher energy at $(\pi,0)$ than at $(\pi/2,\pi/2)$, opposite to the case of weak antiferromagnetic $J_2$ in $\mathrm{Cu(pz)_2(ClO_4)_2}$~\cite{tsyrulin2010two}. Compared to the 39 \% reduction in ordered moment due to quantum fluctuations for the nearest-neighbour Heisenberg model, the estimate for $J_2/J_1=-0.46$ is only a 24 \% reduction of the ordered moment. Adding the weak anisotropy for MoOPO$_4$ yields 21 \% reduction in ordered moment. Hence quantum fluctuations are likely much weaker in MoOPO$_4$ than in e.g. CFTD \cite{ronnow1999high, ronnow2001spin, christensen2007quantum, Piazza15NatP} or Sr$_2$CuTeO$_6$~\cite{Babkevich16PRL}, and it would be interesting in future investigations to examine whether this leads to a similar suppression of the quantum dispersion and continuum around $(\pi,0)$.

\subsection{\emph{Ab initio} calculations}

\begin{figure}[!b]
	\centering
	\includegraphics[width=0.5\textwidth]{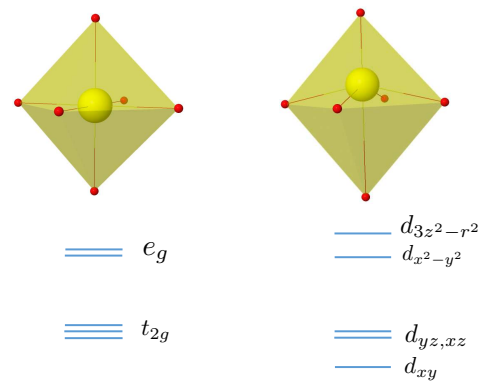}
	\caption{Single-particle energy level diagram of $d$ states in octahedral arrangement of the oxygen ligands (in red) and Mo ion (in yellow) in $\mathrm{MoOPO_4}$.}\label{calc}
\end{figure}

An interesting feature in $\mathrm{MoOPO_4}$ is that the axial position of the Mo$^{4+}$ ion inside MoO$_6$ octahedron is heavily shifted such that the short and long distances to the apical oxygens are 1.652 {\AA} and 2.641 {\AA}, respectively. As a consequence the octahedral symmetry around the Mo ion is reduced, resulting in the removal of octahedral orbital degeneracies and an orbitally mixed electronic ground state. To elucidate the electronic levels of the Mo$^{5+}$ ion in low-symmetry crystal fields in $\mathrm{MoOPO_4}$, we performed \emph{ab~initio} quantum chemistry calculations  using the cluster-in-embedding formalism~\cite{liviu_book}. A cluster of a single active MoO$_6$ octahedron along with surrounding nearest neighbor (NN) PO$_4$ tetrahedra within the plane and the out-of-plane MoO$_6$ octahedra embedded in an array of point charges that reproduces the effect of the solid enviroment~\cite{klintenberg2000120} was considered for many-body calculations. The NN polyhedra were included within the cluster region to better describe the charge density within the active MoO$_6$ region. Such calculations have provided excellent insights into the interplay of crystal field and SOC effects for several $4d$ and $5d$ transition metal compounds~\cite{Katukuri_ioc_Rh,Ir214_katukuri_12,Ir113_bogdanov_12,Ir3116_rixs_liu_2012}.

A perfect octahedral arrangement of the oxygen ligands around the transition-metal ion splits the $d$ levels into high-energy $e_g$ and low-energy $t_{2g}$ manifolds (see Fig.~\ref{calc}). In $\mathrm{MoOPO_4}$, the low-symmetry crystal fields further split the $t_{2g}$ and $e_g$ levels of the Mo$^{5+}$ ion resulting in an orbital singlet ground state. In Table~\ref{dd-calc} the ground-state wavefunction and the $d$-$d$ excitations of the Mo$^{5+}$ ion are summarized. These have been obtained from many-body multi-configurational self consistent field (MCSCF)~\cite{book_QC_00} and $N$-electron valence-state perturbation theory  (NEVPT2)~\cite{Angeli2001297} calculations for the atoms in the active cluster region. All-electron DKH (Douglas-Kroll-Hess) basis sets of triple-zeta quality~\cite{DKH-basis} were used to represent the  Mo and oxygen ions in the central MoO$_6$ octahedron, and for the Mo and P ions in the NN polyhedra we employed effective core potentials~\cite{ECP_stoll_4d_07,ECP_maingroup_Stoll} with valence triple-zeta~\cite{ECP_stoll_4d_07} and a single basis function, respectively. The oxygen ions corresponding to the NN MoO$_6$ and PO$_4$ polyhedra were expanded in atomic natural orbital (ANO) type~\cite{BS_ano_pierloot_95} two $s$ and one $p$ functions. All the caluations were performed using the {\sc orca} quantum chemistry package~\cite{orca}

\begin{table}[!t]
\caption{Relative energies of $d$-level states of Mo$^{5+}$ ion obtained from CASSCF/NEVPT2 calculations. The corresponding wavefunctions without (coefficients) and with (weights) SOC at the CASSCF level are also provided. At NEVPT2 level, the wavefunction would also contain contributions from the inactive and virtual orbitals. For simplicity only the weights of the SOC wavefunction are provided as they are complex.}
\label{dd-calc}
\begin{tabular}{lll}
\hline
\hline\\[-0.25cm]
$t_{2g}^1$ states  &Relative   &Wavefunction (CASSCF) \\
                   &energies (eV) &\\
\hline\\[-0.25cm]
Without SOC\,: & & coefficients\\
$|\phi_0\rangle$  &0           &$0.95\,|xy\rangle - 0.32\,|x^2-y^2\rangle $\\
$|\phi_1\rangle$  &1.79        &$0.98\,|yz\rangle + 0.21\,|zx\rangle$\\
$|\phi_2\rangle$  &1.79        &$0.21\,|yz\rangle - 0.98\,|zx\rangle$\\
$|\phi_3\rangle$  &3.68        &$0.32\,|xy\rangle - 0.95\,|x^2-y^2\rangle $\\
$|\phi_4\rangle$  &4.42        &$1.00\,|z^2\rangle $\\[0.20cm]

With SOC\,:& & normalized weights (\%)\\
$|\psi_0\rangle$  &0           &$\ 86.0\,|\phi_0,\uparrow \rangle +14.0 |\phi_0,\downarrow \rangle$\\
$|\psi_1\rangle$  &1.75        &$\ 50.0\,|\phi_1,\uparrow \rangle + 50.0\,|\phi_2,\downarrow \rangle$\\
$|\psi_2\rangle$  &1.82        &$\ 46.0\,|\phi_1,\uparrow \rangle +46.0 \,|\phi_2,\uparrow \rangle$\\
                  &            &$\ \ 4.0\,|\phi_1,\downarrow \rangle +4.0 \,|\phi_2,\downarrow \rangle$\\
$|\psi_3\rangle$  &3.70        &$\ 88.0\,|\phi_3,\uparrow \rangle + 12.0 \,|\phi_3,\downarrow \rangle$\\
$|\psi_4\rangle$  &4.44        &$100.0\,|\phi_4\rangle$\\
\hline
\hline
\end{tabular}
\end{table}

\begin{table}[!t]
\caption{Computed $g$ factors of MoOPO$_4$ at the NEVPT2 level of theory. The ground state multiconfiguration wavefunction as shown in Table~\ref{dd-calc} produces the correct structure for the $g$ factors.}
\label{table-gs}
\begin{tabular}{c|c|c}
\hline
CASSCF  & $g_a$ & $g_c$ \\
active orbital space& & \\[0.2cm]
\hline
\hline
$t_{2g}$ & 1.91 & 1.99   \\[0.2cm]
$t_{2g}+e_g$ & 1.92 & 1.84 \\[0.2cm]
\hline
\end{tabular}
\end{table}

In the complete active space formalism of the MCSCF (CASSCF) calculation, a self-consistent wavefunction was constructed with an active space of one electron in five Mo $d$ orbitals. On top of the CASSCF wavefunction, NEVPT2 was applied to capture the dynamic electronic correlation. It can be seen in Table~\ref{dd-calc} that the ground state is predominantly of $d_{xy}$ character but has significant contributions from the $d_{x^2-y^2}$ orbital. The first orbital exciations are nearly degenerate at 1.79 eV and are composed of $d_{yz}$ and $d_{zx}$ atomic orbitals. This scenario contrasts the situation in other $t_{2g}$ class of compounds with regular transition-metal oxygen octahedra where the $t_{2g}$ manifold remains degenerate with an effective orbital angular momentum ${\tilde l}=1$. In the latter scenario the spin-orbit interaction admixes all the $t_{2g}$ states to give rise to a total angular momentum $J_\mathrm{eff}$ ground state \cite{Kim09Sci, Kim08PRL}. Due to the large non-cubic crystal field splittings in the $t_{2g}$ manifold in $\mathrm{MoOPO_4}$, the spin-orbit interaction has negligble effect on the Mo$^{5+}$ ground state $\psi_0$, see `with SOC' results in Table \ref{dd-calc}. However, the orbital angular momentum is unquenched in the $d_{zx}$ and $d_{yz}$ and hence the SOC results in the splitting of the high-energy $\psi_1$ and $\psi_2$. Our calculations result in excitation energies of 3.68 eV and 4.42 eV into the $e_g$ states.

To understand the unusual structure of $g$ factors deduced from the ESR experiments, we computed the same  from the {\it ab initio} wavefunction as implemented in {\sc orca}~\cite{orca_gtensor}. In Table~\ref{table-gs}, the $g$ factors obtained from CASSCF calculations with two different active orbital spaces, $t_{2g}$ only and $t_{2g}+e_g$, are presented. With only $t_{2g}$ orbitals in the active space, we find $g_a < g_c$ as expected for tetragonal symmetry with $d_{xy}$ ground state. By enlarging the active space, the multiconfiguration wavefunction now contains configurations involving the $e_g$ orbitals and this is crucial to produce the experimentally observed $g$ factors with $g_a > g_c$.

\section{Conclusion}

We have shown by a variety of experimental and computational techniques that $\mathrm{MoOPO_4}$ realizes a spin-$1/2$ magnetic system of $4d^1$ electrons, with the quenched orbital moment due to the large displacement of the Mo ions inside the MoO$_6$ octahedra. The magnetic ground state supports a N\'eel-type collinear staggered order on the square-lattice with the moments pointing normal to the plane, while the moments align ferromagnetically along the stacking axis. The compound likely realizes a spin-$1/2$ Heisenberg model on a $J_1$-$J_2$ square lattice, with an unfrustrated configuration of antiferromagnetic $J_1$ and ferromagnetic $J_2$. The spin-flop transition suggests a small easy-axis anisotropy in the dominant antiferromagnetic exchange, and the mean-field calculation reproduces the experimental magnetic phase diagram. The small anisotropy in the $g$ factor observed in ESR, which is reproduced by the quantum chemistry calculations, points to that the ground state involves the higher-energy $e_g$ orbitals in addition to the $t_{2g}$ orbitals. Our results suggest that $4d$ molybdates provide an alternative playground to search for model quantum magnets other than $3d$ compounds.

\section*{Acknowledgment}
We thank R. Scopelliti and O. Zaharko for their help with the x-ray and neutron diffraction, respectively. We also thank V. Favre and P. Huang for their help with the specific heat analysis. V.M.K is grateful to H. Stoll for discussions on effective core potentials. This work was supported by the Swiss National Science Foundation, the MPBH network, and European Research Council grants CONQUEST and TopoMat (No. 306504). M.J. is grateful to support by European Commission through Marie Sk{\l}odowska-Curie Action COFUND (EPFL Fellows). M.K. is supported by a Grants-in-Aid for Scientific Research (C) (JSPS, KAKENHI No 15K05140). The {\it ab initio} calculations have been performed at the Swiss National Supercomputing Centre (CSCS) under project s675.  

\bibliography{MoOPO4}

\end{document}